\documentclass[conference,10pt]{IEEEtran}
\ifCLASSOPTIONcompsoc
  \usepackage[nocompress]{cite}
\else
  \usepackage{cite}
\fi
\ifCLASSINFOpdf
\else
\fi

\IEEEoverridecommandlockouts
\usepackage{cite}

\makeatletter
\newcommand{\removelatexerror}{\let\@latex@error\@gobble}
\makeatother

\usepackage{algorithm}
\usepackage[noend]{algpseudocode}

\ifCLASSOPTIONcompsoc
\usepackage[caption=false, font=scriptsize, labelfont=sf, textfont=sf]{subfig}
\else
\usepackage[caption=false, font=footnotesize]{subfig}
\fi

\usepackage{centernot}
\usepackage[pdftex]{graphicx}
\usepackage{balance}
\usepackage{embedfile}
\usepackage{relsize}
\usepackage{epstopdf}
\usepackage{amsmath,amssymb,amsfonts,bm,amsxtra,latexsym,amscd,amsthm,sansmath}
\usepackage{cases}
\usepackage{accents}
\usepackage{soul}
\usepackage[mathscr]{eucal}
\usepackage{graphicx}
\usepackage{textcomp}
\usepackage{xcolor}
\usepackage{makecell}
\usepackage{array}
\usepackage{booktabs} 
\usepackage{multirow} 
\usepackage{diagbox}
\usepackage{mathtools} 
\theoremstyle{definition}

\theoremstyle{plain}

\DeclareMathAlphabet\mathbfcal{OMS}{cmsy}{b}{n}

\makeatother

\usepackage{cuted}
\newcommand{\errsymba}[1]{\accentset{\pmb{\sim}}{#1}}

\DeclareFontFamily{U}{BOONDOX-calo}{\skewchar\font=45 }
\DeclareFontShape{U}{BOONDOX-calo}{m}{n}{
  <-> s*[1.05] BOONDOX-r-calo}{}
\DeclareFontShape{U}{BOONDOX-calo}{b}{n}{
  <-> s*[1.05] BOONDOX-b-calo}{}
\DeclareMathAlphabet{\mathcalboondox}{U}{BOONDOX-calo}{m}{n}
\SetMathAlphabet{\mathcalboondox}{bold}{U}{BOONDOX-calo}{b}{n}
\DeclareMathAlphabet{\mathbcalboondox}{U}{BOONDOX-calo}{b}{n}

\makeatletter
\g@addto@macro\normalsize{%
  \setlength\abovedisplayskip{2.1pt}
  \setlength\belowdisplayskip{2.1pt}
  \setlength\abovedisplayshortskip{2.1pt}
  \setlength\belowdisplayshortskip{2.1pt}
}
\makeatother

\makeatletter
\newcommand\fs@betterruled{%
  \def\@fs@cfont{\bfseries}\let\@fs@capt\floatc@ruled
  \def\@fs@pre{\vspace*{3pt}\hrule height.8pt depth0pt \kern2pt}%
  \def\@fs@post{\kern2pt\hrule\relax}%
  \def\@fs@mid{\kern2pt\hrule\kern2pt}%
  \let\@fs@iftopcapt\iftrue}
\floatstyle{betterruled}
\restylefloat{algorithm}
\makeatother

\begin{document}

\bstctlcite{IEEEexample:BSTcontrol}

\newcommand{\Cbb}{\mathbb{C}}
\newcommand{\Ebb}{\mathbb{E}}
\newcommand{\Nbb}{\mathbb{N}}
\newcommand{\Rbb}{\mathbb{R}}
\newcommand{\AAcalb}{\mathbcalboondox{A}}
\newcommand{\BBcalb}{\mathbcalboondox{B}}
\newcommand{\CCcalb}{\mathbcalboondox{C}}
\newcommand{\DDcalb}{\mathbcalboondox{D}}
\newcommand{\EEcalb}{\mathbcalboondox{E}}
\newcommand{\FFcalb}{\mathbcalboondox{F}}
\newcommand{\GGcalb}{\mathbcalboondox{G}}
\newcommand{\HHcalb}{\mathbcalboondox{H}}
\newcommand{\IIcalb}{\mathbcalboondox{I}}
\newcommand{\JJcalb}{\mathbcalboondox{J}}
\newcommand{\KKcalb}{\mathbcalboondox{K}}
\newcommand{\LLcalb}{\mathbcalboondox{L}}
\newcommand{\MMcalb}{\mathbcalboondox{M}}
\newcommand{\NNcalb}{\mathbcalboondox{N}}
\newcommand{\OOcalb}{\mathbcalboondox{O}}
\newcommand{\PPcalb}{\mathbcalboondox{P}}
\newcommand{\QQcalb}{\mathbcalboondox{Q}}
\newcommand{\RRcalb}{\mathbcalboondox{R}}
\newcommand{\SScalb}{\mathbcalboondox{S}}
\newcommand{\TTcalb}{\mathbcalboondox{T}}
\newcommand{\UUcalb}{\mathbcalboondox{U}}
\newcommand{\VVcalb}{\mathbcalboondox{V}}
\newcommand{\WWcalb}{\mathbcalboondox{W}}
\newcommand{\XXcalb}{\mathbcalboondox{X}}
\newcommand{\ZZcalb}{\mathbcalboondox{Z}}

\newcommand{\acalb}{\mathbcalboondox{a}}
\newcommand{\bcalb}{\mathbcalboondox{b}}
\newcommand{\ccalb}{\mathbcalboondox{c}}
\newcommand{\dcalb}{\mathbcalboondox{d}}
\newcommand{\ecalb}{\mathbcalboondox{e}}
\newcommand{\fcalb}{\mathbcalboondox{f}}
\newcommand{\gcalb}{\mathbcalboondox{g}}
\newcommand{\hcalb}{\mathbcalboondox{h}}
\newcommand{\icalb}{\mathbcalboondox{i}}
\newcommand{\jcalb}{\mathbcalboondox{j}}
\newcommand{\kcalb}{\mathbcalboondox{k}}
\newcommand{\lcalb}{\mathbcalboondox{l}}
\newcommand{\mcalb}{\mathbcalboondox{m}}
\newcommand{\ncalb}{\mathbcalboondox{n}}
\newcommand{\ocalb}{\mathbcalboondox{o}}
\newcommand{\pcalb}{\mathbcalboondox{p}}
\newcommand{\qcalb}{\mathbcalboondox{q}}
\newcommand{\rcalb}{\mathbcalboondox{r}}
\newcommand{\scalb}{\mathbcalboondox{s}}
\newcommand{\tcalb}{\mathbcalboondox{t}}
\newcommand{\ucalb}{\mathbcalboondox{u}}
\newcommand{\vcalb}{\mathbcalboondox{v}}
\newcommand{\wcalb}{\mathbcalboondox{w}}
\newcommand{\xcalb}{\mathbcalboondox{x}}
\newcommand{\zcalb}{\mathbcalboondox{z}}

\newcommand{\Acal}{\mathcal{A}}
\newcommand{\Bcal}{\mathcal{B}}
\newcommand{\Ccal}{\mathcal{C}}
\newcommand{\Dcal}{\mathcal{D}}
\newcommand{\Ecal}{\mathbfcal{E}}
\newcommand{\Fcal}{\mathcal{F}}
\newcommand{\Gcal}{\mathcal{G}}
\newcommand{\Hcal}{\mathcal{H}}
\newcommand{\Ical}{\mathcal{I}}
\newcommand{\Jcal}{\mathcal{J}}
\newcommand{\Kcal}{\mathcal{K}}
\newcommand{\Lcal}{\mathcal{L}}
\newcommand{\Mcal}{\mathcal{M}}
\newcommand{\Ncal}{\mathcal{N}}
\newcommand{\Ocal}{\mathcal{O}}
\newcommand{\Pcal}{\mathcal{P}}
\newcommand{\Qcal}{\mathcal{Q}}
\newcommand{\Rcal}{\mathcal{R}}
\newcommand{\Scal}{\mathcal{S}}
\newcommand{\Tcal}{\mathcal{T}}
\newcommand{\Ucal}{\mathcal{U}}
\newcommand{\Vcal}{\mathcal{V}}
\newcommand{\Wcal}{\mathcal{W}}
\newcommand{\Xcal}{\mathcal{X}}
\newcommand{\Zcal}{\mathcal{Z}}
\newcommand{\Abf}{\boldsymbol{A}}
\newcommand{\Bbf}{\boldsymbol{B}}
\newcommand{\Cbf}{\boldsymbol{C}}
\newcommand{\Dbf}{\boldsymbol{D}}
\newcommand{\Ebf}{\boldsymbol{E}}
\newcommand{\Fbf}{\boldsymbol{F}}
\newcommand{\Gbf}{\boldsymbol{G}}
\newcommand{\Hbf}{\boldsymbol{H}}
\newcommand{\Ibf}{\boldsymbol{I}}
\newcommand{\Jbf}{\boldsymbol{J}}
\newcommand{\Kbf}{\boldsymbol{K}}
\newcommand{\Lbf}{\boldsymbol{L}}
\newcommand{\Mbf}{\boldsymbol{M}}
\newcommand{\Nbf}{\boldsymbol{N}}
\newcommand{\Obf}{\boldsymbol{O}}
\newcommand{\Pbf}{\boldsymbol{P}}
\newcommand{\Qbf}{\boldsymbol{Q}}
\newcommand{\Rbf}{\boldsymbol{R}}
\newcommand{\Sbf}{\boldsymbol{S}}
\newcommand{\Tbf}{\boldsymbol{T}}
\newcommand{\Ubf}{\boldsymbol{U}}
\newcommand{\Vbf}{\boldsymbol{V}}
\newcommand{\Wbf}{\boldsymbol{W}}
\newcommand{\Xbf}{\boldsymbol{X}}
\newcommand{\Zbf}{\boldsymbol{Z}}
\newcommand{\Afrak}{\mathfrak{A}}
\newcommand{\Bfrak}{\mathfrak{B}}
\newcommand{\Cfrak}{\mathfrak{C}}
\newcommand{\Dfrak}{\mathfrak{D}}
\newcommand{\Efrak}{\mathfrak{E}}
\newcommand{\Ffrak}{\mathfrak{F}}
\newcommand{\Gfrak}{\mathfrak{G}}
\newcommand{\Hfrak}{\mathfrak{H}}
\newcommand{\Ifrak}{\mathfrak{I}}
\newcommand{\Jfrak}{\mathfrak{J}}
\newcommand{\Kfrak}{\mathfrak{K}}
\newcommand{\Lfrak}{\mathfrak{L}}
\newcommand{\Mfrak}{\mathfrak{M}}
\newcommand{\Nfrak}{\mathfrak{N}}
\newcommand{\Ofrak}{\mathfrak{O}}
\newcommand{\Pfrak}{\mathfrak{P}}
\newcommand{\Qfrak}{\mathfrak{Q}}
\newcommand{\Rfrak}{\mathfrak{R}}
\newcommand{\Sfrak}{\mathfrak{S}}
\newcommand{\Tfrak}{\mathfrak{T}}
\newcommand{\Ufrak}{\mathfrak{U}}
\newcommand{\Vfrak}{\mathfrak{V}}
\newcommand{\Wfrak}{\mathfrak{W}}
\newcommand{\Xfrak}{\mathfrak{X}}
\newcommand{\Zfrak}{\mathfrak{Z}}

\newcommand{\Ascr}{\mathscr{A}}
\newcommand{\Bscr}{\mathscr{B}}
\newcommand{\Cscr}{\mathscr{C}}
\newcommand{\Dscr}{\mathscr{D}}
\newcommand{\Escr}{\mathscr{E}}
\newcommand{\Fscr}{\mathscr{F}}
\newcommand{\Gscr}{\mathscr{G}}
\newcommand{\Hscr}{\mathscr{H}}
\newcommand{\Iscr}{\mathscr{I}}
\newcommand{\Jscr}{\mathscr{J}}
\newcommand{\Kscr}{\mathscr{K}}
\newcommand{\Lscr}{\mathscr{L}}
\newcommand{\Mscr}{\mathscr{M}}
\newcommand{\Nscr}{\mathscr{N}}
\newcommand{\Oscr}{\mathscr{O}}
\newcommand{\Pscr}{\boldsymbol{\mathscr{P}}}
\newcommand{\Qscr}{\mathscr{Q}}
\newcommand{\Rscr}{\mathscr{R}}
\newcommand{\Sscr}{\mathscr{S}}
\newcommand{\Tscr}{\mathscr{T}}
\newcommand{\Uscr}{\mathscr{U}}
\newcommand{\Vscr}{\mathscr{V}}
\newcommand{\Wscr}{\mathscr{W}}
\newcommand{\Xscr}{\mathscr{X}}
\newcommand{\Zscr}{\mathscr{Z}}

\newcommand{\mub}{\boldsymbol{\mu}}
\newcommand{\m}{\mathrm{m}}
\newcommand{\km}{\mathrm{km}}
\newcommand{\met}{\mathrm{m}}
\newcommand{\mW}{\mathrm{mW}}
\newcommand{\upsib}{\boldsymbol{\upsilon}}
\newcommand{\varsigb}{\boldsymbol{\varsigma}}
\newcommand{\varphiLoS}{\sbar{\boldsymbol{\varphi}}}
\newcommand{\varphiNLoS}{\stilde{\boldsymbol{\varphi}}}

\newcommand{\ReZ}[1]{\mathbf{Re}\left \{#1 \right \}}
\newcommand{\ImZ}[1]{\mathbf{Im}\left \{#1 \right \}}
\newcommand{\CNcal}{\mathcal{N}_{\mathbb{C}}}
\newcommand{\RIS}{\textrm{RIS}}
\newcommand{\PRE}{\textrm{PRE}}
\newcommand{\BS}{\textrm{BS}}
\newcommand{\AP}{\textrm{AP}}
\newcommand{\FAP}{\textrm{FAP}}
\newcommand{\UAV}{\textrm{UAV}}
\newcommand{\user}{\textrm{UE}}
\newcommand{\guser}{\textrm{GU}}
\newcommand{\SN}{\textrm{SN}}
\newcommand{\sns}{\mathrm{s}}

\newcommand{\JCPLth}{\mathrm{JCPL}^{(t)}}
\newcommand{\LSF}{\mathrm{LSF}}
\newcommand{\Watt}{\mathrm{Watt}}
\newcommand{\prob}{\mathrm{Pr}}
\newcommand{\diff}{\mathrm{d}}
\newcommand{\rank}{\mathrm{rank}}
\newcommand{\coher}{\mathrm{c}}
\newcommand{\pil}{\mathrm{p}}
\newcommand{\upl}{\mathrm{ul}}
\newcommand{\diag}[1]{\mathrm{diag}\left [ #1 \right ]}
\newcommand{\sph}{\zeta}
\newcommand{\SINR}{\mathrm{SINR}}
\newcommand{\TN}{\mathrm{TN}}
\newcommand{\Rate}{\mathfrak{R}}
\newcommand{\trace}[1]{\mathrm{tr}\left( #1 \right )}
\newcommand{\tr}[1]{\mathrm{tr}( #1 )}
\newcommand{\nexp}[1]{\mathrm{exp}\left ( #1 \right )}
\newcommand{\arm}{\mathrm{a}}
\newcommand{\drm}{\mathrm{d}}
\newcommand{\dwnl}{\mathrm{dl}}
\newcommand{\crm}{\mathrm{c}}
\newcommand{\DS}{\mathrm{DS}}
\newcommand{\BU}{\mathrm{BU}}
\newcommand{\UI}{\mathrm{UI}}
\newcommand{\SF}{\mathrm{SF}}
\newcommand{\FU}{\mathrm{FU}}
\newcommand{\SU}{\mathrm{SU}}
\newcommand{\PL}{\mathrm{PL}}
\newcommand{\dB}{\mathrm{dB}}
\newcommand{\shad}{\mathrm{shad}}

\newcommand{\modf}{\mathrm{mod}}
\newcommand{\krm}{\mathrm{k}}
\newcommand{\boldk}{\mathbf{k}}
\newcommand{\boldg}{\mathbf{g}}

\newcommand{\Amat}{\mathbf{A}}
\newcommand{\Bmat}{\mathbf{B}}
\newcommand{\Cmat}{\mathbf{C}}
\newcommand{\Dmat}{\mathbf{D}}
\newcommand{\Fmat}{\mathbf{F}}
\newcommand{\Gmat}{\mathbf{G}}
\newcommand{\Noisemat}{\mathbf{\Sigma}}

\newcommand{\Vmat}{\boldsymbol{V}}
\newcommand{\Zmat}{\boldsymbol{Z}}
\newcommand{\Pmat}{\boldsymbol{P}}
\newcommand{\Mmat}{\boldsymbol{M}}

\newcommand{\Dmatbar}{\sbar{\boldsymbol{D}}}
\newcommand{\Amatbar}{\sbar{\boldsymbol{A}}}
\newcommand{\Bmatbar}{\sbar{\boldsymbol{B}}}
\newcommand{\AmatRA}{\Amat^{\mathrm{RA}}}
\newcommand{\onevec}{\boldsymbol{1}}

\newcommand{\EE}{\mathrm{EE}}
\newcommand{\SE}{\mathrm{SE}}
\newcommand{\Ggamma}{\Acute{\Gamma}}
\newcommand{\Gapo}{\Acute{G}}
\newcommand{\Aamat}{\Acute{A}}
\newcommand{\Bbmat}{\Acute{B}}
\newcommand{\Ccmat}{\Acute{C}}
\newcommand{\Ddmat}{\Acute{D}}
\newcommand{\Smat}{\mathbf{S}}
\newcommand{\Wmat}{\mathbf{W}}
\newcommand{\Tmat}{\mathbf{T}}
\newcommand{\Rmat}{\mathbf{R}}
\newcommand{\Qmat}{\mathbf{Q}}
\newcommand{\RHmat}{\mathscr{R}}
\newcommand{\zeromat}{\mathbf{0}}
\newcommand{\Imat}{\mathbf{I}}
\newcommand{\rbold}{\mathbf{r}}
\newcommand{\matvec}{\mathbf{vec}}
\newcommand{\newid}{\xi}

\newcommand{\NbU}{\boldsymbol{NB}^\user}
\newcommand{\NbA}{\boldsymbol{NB}^\AP}
\newcommand{\sbarNbA}{\sbar{\boldsymbol{NB}}^\AP}
\newcommand{\sumSt}[1]{\sum\limits_{#1\in \Scal_{\varphib(k)}}}
\newcommand{\sumStk}[2]{\sum\limits_{\substack{#1\in \Scal_{\varphib(k)} \\ #1\ne #2}}}
\newcommand{\sumStkk}[2]{\sum\limits_{#1\in \Scal_{\varphib_k}, #1\ne #2}}

\newcommand\ddfrac[2]{\left ( {#1} \right ) / \left ( {#2} \right )}

\newcommand{\astdot}[1]{\accentset{*}{#1}}
\newcommand{\sqrmat}[1]{\accentset{\diamond}{#1}}
\newcommand{\circhat}[1]{\accentset{\circ}{#1}}

\newcommand{\RadiU}{R^\user}
\newcommand{\RadiA}{R^\AP}
\newcommand{\NewUEs}{\check{\mathcal{K}}}
\newcommand{\card}[1]{\left | #1 \right |}
\newcommand{\checkS}{\check{\mathscr{S}}}

\newcommand{\intern}{\mathrm{int}}
\newcommand{\ext}{\mathrm{ext}}
\newcommand{\MsSet}{\boldsymbol{Mst}}
\newcommand{\MemSet}{\mathcal{M}}
\newcommand{\sbarMemSet}{\sbar{\MemSet}}

\newcommand{\TX}{\mathrm{Tx}}
\newcommand{\RX}{\mathrm{Rx}}

\newcommand{\arr}{\boldsymbol{a}}
\newcommand{\az}{\mathrm{az}}
\newcommand{\ele}{\mathrm{ele}}

\newcommand{\aval}{\boldsymbol{a}}
\newcommand{\bval}{\boldsymbol{b}}
\newcommand{\cval}{\boldsymbol{c}}
\newcommand{\dval}{\boldsymbol{d}}
\newcommand{\bvalbar}{\sbar{\boldsymbol{b}}}
\newcommand{\cvalbar}{\sbar{\boldsymbol{c}}}

\newcommand{\Eve}{\textrm{Eve}}
\newcommand{\Erm}{\mathrm{E}}
\newcommand{\APJam}{\textrm{AP-Jammer}}
\newcommand{\Pow}{\boldsymbol{P}}

\newcommand\dsum[2]{{\displaystyle\sum_{#1}^{#2}}}
\newcommand{\lowround}[1]{\left \lfloor #1 \right \rfloor}

\newcommand{\AoA}{\textrm{AoA}}
\newcommand{\AoD}{\textrm{AoD}}

\newcommand{\MMF}{\mathrm{MMF}}
\newcommand{\chg}{\mathbf{g}}
\newcommand{\chh}{\mathbf{h}}
\newcommand{\chF}{\mathbf{F}}
\newcommand{\chf}{\mathbf{f}}
\newcommand{\chG}{\mathbf{G}}
\newcommand{\chH}{\mathbf{H}}
\newcommand{\chGG}{\boldsymbolcal{G}}
\newcommand{\chFF}{\boldsymbolcal{F}}
\newcommand{\errFF}{\errsymba{\boldsymbolcal{F}}}
\newcommand{\errG}{\errsymba{\boldsymbolcal{F}}}
\newcommand{\kmsq}{\mathrm{km}^2}

\newcommand{\chkylk}{\check{y}_{p,lk}}
\newcommand{\chkylkstar}{\check{y}^{*}_{p,lk}}

\newcommand{\chkykl}{\check{y}_{p,kl}}
\newcommand{\chkyklstar}{\check{y}^{*}_{p,kl}}

\newcommand{\cov}[1]{\boldsymbol{cov}\left \{ #1 \right \}}
\newcommand{\mean}[1]{\mathbb{E}\left \{ #1 \right \}}
\newcommand{\variance}[1]{\mathrm{Var}\left \{ #1 \right \}}

\newcommand{\vect}[1]{\mathrm{vec}\left ( #1 \right )}

\newcommand{\matele}[1]{\left [ #1 \right ]}
\newcommand{\groupele}[1]{\left ( #1 \right )}
\newcommand{\vecnorm}[1]{\lVert #1 \rVert}
\newcommand{\vecnormstar}[1]{\left \| #1 \right \|_{\star}}
\newcommand{\numabs}[1]{\lvert #1 \rvert}

\newcommand{\mat}[1]{\boldsymbol{Mat}\left ( #1 \right )}


\newcommand{\src}{\boldsymbol{(S)}}
\newcommand{\dst}{\boldsymbol{(D)}}
\newcommand{\lastvisit}{\boldsymbol{LV}}
\newcommand{\routecost}{\boldsymbol{RC}}
\newcommand{\rt}{\mathrm{rt}}
\newcommand{\DSet}{\mathscr{D}}
\newcommand{\NB}{\boldsymbol{NB}}
\newcommand{\thres}{\mathrm{th}}
\newcommand{\nextvisit}{\boldsymbol{NV}}
\newcommand{\CL}{\boldsymbol{CL}}
\newcommand{\cost}{\boldsymbol{cost}}
\newcommand{\rreq}{\mathrm{rreq}}
\newcommand{\hopcount}{\boldsymbol{hc}}
\newcommand{\NextSet}{\boldsymbol{NH}}
\newcommand{\tot}{\text{tot}}
\newcommand{\dsp}{\text{dsp}}
\newcommand{\hov}{\text{hov}}

\newcommand{\mth}[1]{#1\mathrm{-th}}
\newcommand{\re}{\mathrm{re}}

\newcommand\sbar[1]{\accentset{\rule{.6em}{.8pt}}{#1}}
\newcommand\stilde[1]{\accentset{\sim}{#1}}

\newcommand{\pw}{p}
\newcommand{\pww}{\eta}
\newcommand{\pwwb}{\boldsymbol{\eta}}
\newcommand{\omegab}{\boldsymbol{\omega}}

\newcommand{\boldXX}{\mathbf{X}}
\newcommand{\boldYY}{\mathbf{Y}}
\newcommand{\boldUU}{\mathbf{U}}
\newcommand{\boldZZ}{\mathbf{Z}}
\newcommand{\boldZZA}{\mathbf{\circhat{Z}}}
\newcommand{\checkZ}{\check{Z}}
\newcommand{\circZ}{\circhat{Z}}

\newcommand{\boldh}{\mathbf{h}}
\newcommand{\boldu}{\mathbf{u}}
\newcommand{\boldw}{\mathbf{w}}
\newcommand{\bolda}{\mathbf{a}}
\newcommand{\boldr}{\mathbf{r}}
\newcommand{\boldf}{\mathbf{f}}
\newcommand{\gammab}{\boldsymbol{\gamma}}
\newcommand{\xib}{\boldsymbol{\xi}}

\newcommand{\rhob}{\boldsymbol{\rho}}
\newcommand{\varrhob}{\boldsymbol{\varrho}}

\newcommand{\boldx}{\mathbf{x}}
\newcommand{\boldp}{\mathbf{p}}
\newcommand{\boldb}{\mathbf{b}}
\newcommand{\boldc}{\mathbf{c}}
\newcommand{\boldz}{\mathbf{z}}
\newcommand{\boldy}{\mathbf{y}}
\newcommand{\bnu}{\boldsymbol{\nu}}
\newcommand{\deltab}{\boldsymbol{\delta}}
\newcommand{\boldn}{\mathbf{n}}
\newcommand{\psib}{\boldsymbol{\psi}}
\newcommand{\phib}{\boldsymbol{\phi}}
\newcommand{\varphib}{\boldsymbol{\varphi}}
\newcommand{\vrhob}{\boldsymbol{\varrho}}
\newcommand{\PPhib}{\boldsymbol{\Phi}}

\newcommand{\Sigmab}{\boldsymbol{\Sigma}}
\newcommand{\Dec}{\mathrm{Decode}}
\newcommand{\Enc}{\mathrm{Encode}}

\newcommand{\ZZcal}{\boldsymbolcal{Z}}
\newcommand{\boldv}{\mathbf{v}}
\newcommand{\boldd}{\mathbf{d}}
\newcommand{\boldVV}{\mathbf{V}}
\newcommand{\UESet}{\mathcal{K}}
\newcommand{\GUSet}{\mathcal{K}}

\newcommand{\FAPSet}{\mathcal{L}}
\newcommand{\RISSet}{\mathcal{R}}
\newcommand{\GGcal}{\boldsymbolcal{G}}
\newcommand{\AAcal}{\boldsymbolcal{A}}
\newcommand{\CCcal}{\boldsymbolcal{C}}
\newcommand{\lRk}{l{\RISSet}k}
\newcommand{\lRi}{l{\RISSet}i}
\newcommand{\dps}{\displaystyle}
\newcommand{\MMSE}{\mathrm{MMSE}}
\newcommand{\LSFD}{\mathrm{LSFD}}
\newcommand{\PLSFD}{\mathrm{P-LSFD}}
\newcommand{\PMMSE}{\mathrm{P-MMSE}}
\newcommand{\LPMMSE}{\mathrm{LP-MMSE}}
\newcommand{\opt}{\textrm{opt}}
\newcommand{\popt}{\textrm{p-opt}}
\newcommand{\Bscrb}{\boldsymbol{\Bscr}}

\newcommand{\LoS}{\mathrm{LoS}}
\newcommand{\NLoS}{\mathrm{NLoS}}
\newcommand{\aLoS}{\sbar{\alpha}}
\newcommand{\aNLoS}{\stilde{\alpha}}
\newcommand{\bLoS}{\sbar{\beta}}
\newcommand{\bNLoS}{\stilde{\beta}}

\newcommand{\phiLoS}{\sbar{\phi}}
\newcommand{\phiNLoS}{\stilde{\phi}}

\newcommand{\FFLoS}{\sbar{\boldsymbol{F}}}
\newcommand{\FFNLoS}{\stilde{\boldsymbol{F}}}

\newcommand{\GGLoS}{\sbar{\boldsymbol{G}}}
\newcommand{\GGNLoS}{\stilde{\boldsymbol{G}}}

\newcommand{\gLoS}{\sbar{\boldsymbol{g}}}
\newcommand{\gNLoS}{\stilde{\boldsymbol{g}}}

\newcommand{\fLoS}{\sbar{\boldsymbol{f}}}
\newcommand{\fNLoS}{\stilde{\boldsymbol{f}}}

\newcommand{\frmLoS}{\sbar{f}}
\newcommand{\frmNLoS}{\stilde{f}}

\newcommand{\hLoS}{\sbar{\boldsymbol{h}}}
\newcommand{\hNLoS}{\stilde{\boldsymbol{h}}}

\newcommand{\GlrLoS}{\GGLoS_{lr}}
\newcommand{\GlrNLoS}{\GGNLoS_{lr}}

\newcommand{\Trm}{\mathrm{T}}
\newcommand{\Hrm}{\mathrm{H}}
\newcommand{\Vrm}{\mathrm{V}}
\newcommand{\Urm}{\mathrm{U}}
\newcommand{\Nrm}{\mathrm{N}}
\newcommand{\Frm}{\mathrm{F}}
\newcommand{\Srm}{\mathrm{S}}

\newcommand{\Pcalbar}{\sbar{\mathcal{P}}}

\title{Energy Efficiency Optimization in \\Integrated Satellite-Terrestrial UAV-Enabled \\Cell-Free Massive MIMO
\thanks{This work was funded by the European Union under the Italian National Recovery and Resilience Plan (NRRP) of NextGenerationEU through the PRIN 2022 project ``RAIN4C - Reliable aerial and satellite networks: Joint communication, computation, caching for critical scenarios'' (CUP: J53D23007020001, id: 20227N3SPN) by the Italian MUR.}
}

\author{%
    \IEEEauthorblockN{%
        Thong-Nhat Tran\IEEEauthorrefmark{1},
        Giovanni Interdonato\IEEEauthorrefmark{2}\IEEEauthorrefmark{3}        
    }%
    \IEEEauthorblockA{%
        \IEEEauthorrefmark{1}\textit{Dept. of Information and Communication Engineering, Chungbuk National University, 28644 Cheongju, South Korea.}\\
        \IEEEauthorrefmark{2}\textit{Dept. of Electrical and Information Engineering, University of Cassino and Southern Lazio, 03043 Cassino, Italy.}\\
        \IEEEauthorrefmark{3}\textit{Consorzio Nazionale Interuniversitario per le Telecomunicazioni (CNIT), 43124 Parma, Italy.}\\    
        \texttt{trannhat@gmail.com, giovanni.interdonato@unicas.it}
    }
}

\begin{figure*}[t!]
\normalsize
This paper has been submitted for publication in the proceedings of an IEEE conference.

\

\textcopyright~2024 IEEE. Personal use of this material is permitted. 
Permission from IEEE must be obtained for all other uses, in any current or future media, including reprinting/republishing this material for advertising or promotional purposes, creating new collective works, for resale or redistribution to servers or lists, or reuse of any copyrighted component of this work in other works.

\


\vspace{17cm}
\end{figure*}

\maketitle

\begin{abstract}
    Integrating cell-free massive MIMO (CF-mMIMO) into satellite-unmanned aerial vehicle (UAV) networks offers an effective solution for enhancing connectivity. In this setup, UAVs  serve as access points (APs) of a terrestrial CF-mMIMO network extending the satellite network capabilities, thereby ensuring robust, high-quality communication links. In this work, we propose a successive convex approximation algorithm for maximizing the downlink energy efficiency (EE) at the UAVs under per-UAV power budget and user quality-of-service constraints. We derive a closed-form expression for the EE that accounts for maximum-ratio transmission and statistical channel knowledge at the users. Simulation results show the effectiveness of the proposed algorithm in maximizing the EE at the UAV layer. Moreover, we observe that a few tens of UAVs transmitting with a fine-tuned power are sufficient to empower the service of satellite networks and significantly increase the spectral efficiency.  
\end{abstract}


\section{Introduction}
The integration of satellite and unmanned aerial vehicle (UAV) networks emerges as an effective solution to enhance communication infrastructures, from the remoteness of rural areas to the density of urban centers, leveraging their combined strengths, namely global connectivity and adaptable coverage, respectively. \cite{Azari9861699} provides a survey on the evolution of non-terrestrial networks (NTNs), focusing on their integration with terrestrial networks (TNs) and discussing challenges and future promising prospects, including the satellites-UAVs integration. Notably, the effectiveness of this integration hinges on the massive multiple-input multiple-output (MIMO) physical layer technology~\cite{You9110855,Garcia8869706,van2024joint}.

Cell-free massive MIMO (CF-mMIMO)~\cite{ngo2024ultra,Demir9336188} is deemed to be a key physical layer technology for 6G systems. It combines the intrinsic efficiency, practicality and scalability of the massive MIMO operation with a ultra-dense distributed deployment and the joint coherent multipoint processing. In this regard, implementing CF-mMIMO into satellite-UAV networks offers an attractive and effective solution for enhancing connectivity and achieving unprecedented practical gains~\cite{Vilor9930941, Liu9174846, Lee9944375, Wu10002346}.
In this setup, UAVs serve as access points (APs) operating in a CF-mMIMO setup and extending satellite network capabilities to ensure robust, high-quality communication links. This fusion addresses the limitations of direct satellite communication by providing adaptable coverage and uniform service quality across challenging terrains and environments. Consequently, the satellite-UAV integration employing CF-mMIMO architecture at the UAV-layer opens to new avenues for reliable, scalable, and efficient global connectivity, especially valuable for the Internet of Things (IoT), emergency services, and rural communications.
Specifically, in \cite{Liu9174846}, the authors integrated UAVs with satellites in a cell-free satellite-UAV network to improve 6G IoT connectivity. They proposed a multi-domain resource allocation framework that optimizes subchannels allocation, transmit power, and UAV hovering times, thereby improving network efficiency and ensuring fairness among users. In \cite{Lee9944375}, integrating LEO satellites and UAVs with multi-agent deep reinforcement learning was shown to optimize packet forwarding in hybrid free-space optics/radio frequency (FSO/RF) NTNs, leading to significant improvements in throughput and energy efficiency. In \cite{Wu10002346}, the authors proposed an optimization framework for trajectory and power allocation in a cell-free cognitive satellite-UAV network for IoT, employing mixed-integer non-convex optimization, particle swarm optimization for trajectory, and novel algorithms for power allocation, showcasing enhanced NTN communication.
However, to the best of authors' knowledge, the energy efficiency issues at the UAVs in an integrated satellite-terrestrial UAV-enabled cell-free network remain unexplored.

\textbf{Contributions:} We investigate an integrated satellite-terrestrial UAV-enabled CF-mMIMO system wherein UAVs serve as APs to enhance connectivity in an area covered by satellite communications. Our focus is aiding the satellite communications with a terrestrial UAV-enabled cell-free network to provide adaptable coverage and uniformly great quality of service (QoS).
In this regard, we propose a successive convex approximation (SCA) algorithm to maximize the downlink (DL) energy efficiency (EE) of the UAV layer of the integrated satellite-terrestrial UAV-enabled cell-free network, subject to per-UAV power budget constraints and user's QoS requirements. We derive a closed-form expression for the DL EE that accounts for maximum-ratio (MR) transmission and statistical channel knowledge at the ground users.

\section{System Model}
We consider an integrated satellite-terrestrial UAV-enabled CF-mMIMO network designed to enhance connectivity in critical scenarios, such as remote or rural areas where laying traditional infrastructure is challenging or costly, or a post-disaster scenario where terrestrial infrastructure and fiber links may be temporary lacking or out-of-service. 
The considered system operates in time division duplex (TDD) mode and consists of a low earth orbit (LEO) satellite and $L$ UAVs jointly and coherently serving $K$ single-antenna ground users (GUs). The satellite, equipped with a uniform linear array (ULA) with $N$ antennas (typically ranging from 100 to 196), serves as the central node for broad area coverage and coordination of UAVs. Each UAV is equipped with a ULA with $M$ antennas (typically small, up to 8 antenna elements) and serves as aerial AP of a terrestrial cell-free network. We assume that the UAVs are deployed in a uniform grid pattern and hover the coverage area. 
Both the satellite and the UAVs operate in the same sub-6GHz frequency band to provide data service to the GUs in a cell-free operation. 
We assume that payload data is readily available at the satellite and UAVs in the DL. 
Our aim is maximizing the DL EE of the UAV layer of the integrated satellite-UAVs CF-mMIMO system with respect to the DL transmit powers. Clearly, the transmit power has a negligible impact on the overall energy consumption budget at the satellite. Hence, we only focus on maximizing the EE with respect to the UAV transmit powers, which play, on the other hand, a crucial role in the energy budget of such battery-powered devices.

\subsection{Channel Model}
We assume perfect channel state information (CSI) at both UAVs and satellite, while GUs have access only to statistical channel information. 
We assume spatially correlated Rician fading channels~\cite{You8698468} both for the TN and the NTN. Let $\chh_{l,k}$ denote the channel between an arbitrary $\guser_k$ and $\AP_l$\footnote{In this paper, we use UAV and AP as interchangeable terms.}, then
\begin{equation}
\chh_{l,k} = \sbar{\chh}_{l,k} + \stilde{\chh}_{l,k} \in \Cbb^M,
\end{equation}
where $\sbar{\chh}_{l,k}$ is the line-of-sight (LoS) component, $\stilde{\chh}_{l,k}\!\sim\! \CNcal(0,\Rmat^{\chh}_{l,k})$ and $\Rmat^{\chh}_{l,k}\!\in\!\Cbb^{M\times M}$ is a positive semi-definite covariance matrix that encompasses the spatial correlation of the non-LoS (NLoS) component. We assume the standard block fading model, according to which the channels stay constant within the coherence block and varies independently over successive blocks. The LoS component of $\chh_{l,k}$ is
\begin{equation}
    \sbar{\chh}_{l,k} = \sqrt{\sbar{\prob}^{\chh}_{l,k} \sbar{\beta}^{\chh}_{l,k} \dfrac{\kappa^{\chh}_{l,k}}{\kappa^{\chh}_{l,k}+1}} \bolda^{\chh}_{l,k}(\varphi) \in \Cbb^M,
\end{equation}
where the large-scale fading (LSF) coefficient of the LoS link is $\sbar{\beta}^{\chh}_{l,k}$, $\kappa^{\chh}_{l,k}$ represents the Rician factor, $\sbar\prob^{\chh}_{l,k}$ is the probability of having a LoS link, and $\bolda^{\chh}_{l,k}$ is the array's response corresponding to channel $\chh_{l,k}.$ Assuming the Gaussian local scattering model \cite{Demir9336188}, we determine the $\mth{(m,n)}$ element of the Toeplitz matrix $\mathbf{R}^{\chh}_{l,k}$ through the expression
\begin{equation}
    \matele{\Rmat^{\chh}_{l,k}}_{m,n} = \dfrac{\stilde{\Pr}^{\chh}_{l,k} \stilde{\beta}^{\chh}_{l,k}}{\kappa^{\chh}_{l,k} + 1} \int_{-\infty}^{+\infty} e^{j\pi (m-n)\sin(\varphi+\delta)} f_{\delta} (\delta) \diff \delta,
\end{equation}
wherein $\stilde{\beta}^{\chh}_{l,k}$ is the LSF coefficient of the NLoS link, $\stilde\prob^{\chh}_{l,k}$ is the probability of having a NLoS link, $f_{\delta}(\delta)$ is the probability density function (PDF) reflecting the variability in angular deviations encountered by the propagation paths due to local scattering in proximity to the GU relative to the nominal angle-of-arrival (AoA). We assume the angular deviation follows a Gaussian distribution $\delta \!\sim\! \Ncal(0, \sigma^2_{\delta})$, where $\sigma_{\delta}$ is the angular standard deviation (ASD) around the nominal AoA. We let 
\begin{equation}\label{mean_h}
    \begin{array}{l}
        \mub^{\chh}_{l,k} \triangleq \mean{\chh_{l,k}} = \sbar{\chh}_{l,k},  \\
        \Ecal^{\chh}_{l,k} \triangleq \mean{\chh_{l,k} \chh^\Hrm_{l,k}} = \sbar{\chh}_{l,k} \sbar{\chh}^\Hrm_{l,k} + \Rmat^{\chh}_{l,k}\,,
    \end{array}
\end{equation}
be the mean and the spatial correlation matrix of the terrestrial channel between the $k$-th GU and the $l$-th AP.
The non-terrestrial channel between the satellite and an arbitrary $\guser_k$~is 
\begin{equation}
\chg_{k} = \sbar{\chg}_k + \stilde{\chg}_{k} \in \Cbb^N,
\end{equation}
where $\sbar{\chg}_{k}$ is the LoS component, $\stilde{\chg}_k\sim \CNcal(0,\Rmat^{\chg}_{k})$ and $\Rmat^{\chg}_k$ is a positive semi-definite covariance matrix that encompasses the spatial correlation of the NLoS component. We let 
\begin{equation}\label{mean_g}
    \begin{array}{l}
        \mub^{\chg}_k \triangleq \mean{\chg_k} = \sbar{\chg}_k,  \\
        \Ecal^{\chg}_k \triangleq \mean{\chg_k \chg^\Hrm_k} = \sbar{\chg}_k \sbar{\chg}^\Hrm_k + \Rmat^{\chg}_k\, ,
    \end{array}
\end{equation}
be the mean and the spatial correlation matrix of the non-terrestrial channel between the $k$-th GU and the satellite.

\subsection{DL Spectral and Energy Efficiency}
In the DL data transmission phase, the satellite and UAVs simultaneously transmit data to all the GUs. Let $s_i \!\in\! \Cbb$ be the data symbol for $\guser_{i},$ with $\mean{|s_i|^2} \!=\! 1$ and $\mean{s_is^*_k}\!=\!0,\, \forall i \ne k$. Let us denote the local precoding vector designed by $\AP_l$ and intended for $\guser_i$ by $\boldw^{\AP}_{l,i}\in \Cbb^M.$ The data signal transmitted by $\AP_l$ is given by
\begin{equation}
    \boldx^{\AP}_l = \sum_{i=1}^K \sqrt{\pww^{\AP}_{l,i}} \boldw^{\AP}_{l,i} s_i \in \Cbb^M,
\end{equation}
where $\pww^{\AP}_{l,i}$ represents the power control coefficients of the link $\AP_l$-to-$\guser_i$ that satisfies the power constraint $\mean{\vecnorm{\boldx^{\AP}_l}^2} \!\leq\! P^{\AP}_{\dwnl},$ and $P^{\AP}_{\dwnl}$ is the maximum transmit power at $\AP_l$. Similarly, the data signal transmitted by the satellite (SN) is
\begin{equation}
   \boldx^{\SN} = \sum_{i=1}^K \sqrt{\pww^{\SN}_i} \boldw^{\SN}_i s_i \in \Cbb^N,
\end{equation}
where $\boldw^{\SN}_{i} \!\in\! \Cbb^N$ is the precoding vector at the SN intended for $\guser_i$, while $\pww^{\SN}_i$ is the power control coefficient set to satisfy $\mean{\vecnorm{\boldx^{\SN}}^2} \!\leq\! P^{\SN}_{\dwnl}.$
The received signal at $\guser_k$ is
\begin{equation} \label{eq:received-signal}
    \arraycolsep=2.5pt\def\arraystretch{1.5}
    \begin{array}{ll}
        y^{\dwnl}_k = & \sum\limits_{i=1}^K \sqrt{\pww^{\SN}} \chg^\Hrm_k \boldw^{\SN}_i s_i + \sum
        \limits_{l=1}^L \sqrt{\pww^{\AP}_{l,k}} \chh^\Hrm_{l,k} \boldw^{\AP}_{l,k} s_k \\
        & + \sum\limits_{\{i=1, i\ne k\}}^K \sum\limits_{l=1}^L  \sqrt{\pww^{\AP}_{l,i}} \chh^\Hrm_{l,k} \boldw^{\AP}_{l,i} s_i + n_k \in \Cbb,
    \end{array}
\end{equation}
where $n_{k} \!\sim\! \CNcal(0, \sigma^2_{\dwnl})$ is the receiver noise, which is identically independent distributed with variance $\sigma^2_{\dwnl}$. 
Let us define $\pwwb\!\triangleq\!\{\pwwb_i\}^K_{i=1}$, with $\pwwb_k \!\triangleq\! [\sqrt{\pww_{1,k}}, \ldots, \sqrt{\pww_{L,k}}]^\Trm$, $\boldb_{k,i} \!\triangleq\! \mean{\psib_{k,i}}$, with
$\psib_{k,i} \!\triangleq\! (\chh^\Hrm_{1,k} \boldw^{\AP}_{1,i}, \ldots, \chh^\Hrm_{L,k} \boldw^{\AP}_{L,i})^\Hrm,$ and 
\begin{equation*}
\Cmat^2_{k,i} \!\triangleq\!
    \begin{cases}
        \mean{\psib_{k,i}\psib^\Hrm_{k,i}}, &\text{if } i \!\ne\! k, \\
        \mean{\psib_{k,i}\psib^\Hrm_{k,i}} \!-\! \mean{\psib_{k,i}}\mean{\psib^\Hrm_{k,i}}, &\text{otherwise.}
    \end{cases}    
\end{equation*}
An achievable DL spectral efficiency (SE) can be obtained by applying the \textit{hardening bound} technique~\cite{Demir9336188} on~\eqref{eq:received-signal}, by treating all the interference sources as uncorrelated noise. Specifically, an achievable SE of an arbitrary $\guser_k$ is given by 
\begin{equation}
    \SE^{(\dwnl)}_k (\pwwb) = \log_2 \big(1 + \SINR^{(\dwnl)}_k (\pwwb)\big),
\end{equation}
where the signal-to-interference-plus-noise ratio (SINR) is
\begin{equation}\label{sinr_dwnl}
    \begin{array}{l}
        \SINR^{(\dwnl)}_k (\pwwb) = \dfrac{\pww^{\SN}_k\numabs{\mean{\chg^\Hrm_k \boldw^\SN_k}}^2 + \numabs{\boldb^\Hrm_{k,k} \pwwb_k}^2}{B_k + \sum\limits_{i=1}^K \vecnorm{\Cmat^\Hrm_{k,i}\pwwb_i}^2 + \sigma^2_{\dwnl}}, 
    \end{array}
\end{equation}
where $B_k \triangleq \sum\limits_{i=1}^K \pww^{\SN}_i \mean{\numabs{\chg^\Hrm_k\boldw^{\SN}_i}^2} - \pww^{\SN}_k\numabs{\mean{\chg^\Hrm_{k}\boldw^{\SN}_k}}^2$.
The total power consumption at the UAVs is given by
\begin{equation}
    P_{\tot} \!=\! \sum\limits^{L}_{l = 1} (P_{l,\text{tx}} \!+\! P_{l,0}) \!=\! \sum\limits^{L}_{l = 1} \left(\frac{1}{\epsilon_l} \mean{\lVert{\boldx^{\AP}_l}\rVert^2} \!+\! P_{l,0}\right),
\end{equation}
where $\epsilon_l \!\in\! [0,1]$ is the power amplifier efficiency and $P_{l,0} \!\triangleq\! M P_{l,\dsp} \!+\! P_{l,\hov}$, consists of the power consumed for signal processing at each antenna of the $l$-th UAV and the power cost for hovering.
The total EE at the UAVs is thus given by
\begin{equation} \label{eq:energy-efficiency}
    \!\!\EE(\pwwb) \!=\! \frac{\sum\nolimits_{k=1}^{K} \SE^{(\dwnl)}_k (\pwwb)}{\dfrac{1}{\epsilon}\sum\limits^{L}_{l = 1} \sum\limits^K_{i = 1} \pww^{\AP}_{l,i} \,\mean{\lVert\boldw^{\AP}_{l,i}\rVert^2} \!+\! L M P_{\dsp} \!+\! LP_{\hov}},
\end{equation}
assuming, without loss of generality, that the UAVs are equipped with the same hardware.
MR is computationally the cheapest precoding scheme allowing for a fully distributed implementation. It consists in setting $\boldw^{\AP}_{l,k} \!=\! \chh_{l,k}$, $\forall k,\, \forall l$, and $\boldw^{\SN}_k \!=\! \chg_k$.
Importantly, MR enables us to compute the SE, and in turn the EE, in closed form by properly replacing the following results into the SINR expression in~\eqref{sinr_dwnl}:
\begin{align}
    &\matele{\mean{\psib_{k,k}\psib^\Hrm_{k,k}}}_{l,l} \!=\! \mean{\vecnorm{\chh_{l,k}}^4} \!=\!\sum\limits_{m=1}^{M} \mean{\numabs{\matele{\chh_{l,k}}_m}^4} \nonumber \\[-0.5em]
    &\qquad \!+\! \sum\limits_{\vphantom{m'\ne m}m=1}^{M}\sum\limits_{m'\ne m}^{M} \mean{\numabs{\matele{\chh_{l,k}}_m}^2} \mean{\numabs{\matele{\chh_{l,k}}_{m'}}^2}\, ,
\end{align}
\begin{align}
    &\matele{\mean{\psib_{k,k}}}_l \!=\! \mean{\chh^{\Hrm}_{l,k} \chh_{l,k}} = \trace{\Ecal^{\chh}_{l,k}}, \\
    &\matele{\mean{\psib_{k,i}}}_l \!=\! \mean{\chh^{\Hrm}_{l,i} \chh_{l,k}} = \groupele{\mub^{\chh}_{l,i}}^\Hrm \mub^{\chh}_{l,k}\,,\, \forall i\!\ne\! k,  \\
    &\matele{\mean{\psib_{k,k}\psib^\Hrm_{k,k}}}_{l,l'} \!=\!  \trace{\Ecal^{\chh}_{l,k}} \trace{\Ecal^{\chh}_{l',k}}, \, \forall l'\!\ne\! l, \\
    &\matele{\mean{\psib_{k,i}\psib^\Hrm_{k,i}}}_{l,l} \!=\!  \trace{\Ecal^{\chh}_{l,i} \Ecal^{\chh}_{l,k}},\, \forall i\!\ne\! k, \\
    &\matele{\mean{\psib_{k,i}\psib^\Hrm_{k,i}}}_{l,l'} \!=\!   \groupele{\mub^{\chh}_{l,i}}^{\!\Hrm}\!\! \mub^{\chh}_{l,k} \groupele{\mub^{\chh}_{l',k}}^{\!\Hrm}\!\!\mub^{\chh}_{l',i}, \, \forall i\!\ne\! k, \forall l'\!\ne\! l.
\end{align}
Similarly, as for the link from the satellite to $\guser_k$, we have 
\begin{align}
    \mean{\chg^\Hrm_k \boldw^\SN_k} &=\! \trace{\Ecal^{\chg}_k}\,, \\
    \mean{\numabs{\chg^\Hrm_k\boldw^{\SN}_k}^2} &=\! \mean{\vecnorm{\chg_k}^4}.
\end{align}

\section{Energy Efficiency Maximization}
The EE maximization problem (EEM) is formulated as
\begin{subequations}\label{mee1}
\begin{align}
    \max_{\pwwb} &\; \EE(\pwwb) \tag{\ref{mee1}} \\[-.4em]
    \text{s.t.} &\; \sum\limits_{i=1}^K \pww^{\AP}_{l,i} \, \mean{\vecnorm{\boldw^{\AP}_{l,i}}^2} \!\leq\! P^{\AP}_{\dwnl}, \, \forall l \!=\! 1, \ldots, L, \label{mee1_a} \\
    &\; \SE^{(\dwnl)}_k (\pwwb) \geq \SE^{\min}_k, \, \forall k \!=\! 1, \ldots, K, \label{mee1_b} 
\end{align}
\end{subequations}
where $\SE^{\min}_k$ is a minimum SE threshold for GU $k$.
Firstly, we introduce the auxiliary variables $\boldr \!=\! \{r_i\}_{i=1}^K,$ $\gammab \!=\! \{\gamma_i\}_{i=1}^K$ and $t$ to reformulate~\eqref{mee1} as
\begin{subequations}\label{mee2}
\begin{align}
    \max_{\substack{\pwwb,\, \boldr,\\ \gammab,\, t}} & \sum\nolimits_{k=1}^K r_k \tag{\ref{mee2}} \\
    \text{s.t. } & \eqref{mee1_a} \\[-.4em]
    & \eqref{mee1_b} \Leftrightarrow \log_2 \groupele{1+1/\gamma_k} \geq \SE^{\min}_k, \\[-.4em]
    & \dfrac{1}{\epsilon}\!\sum\limits^{L}_{l = 1} \!\sum\limits^K_{i = 1} \!\pww^{\AP}_{l,i} \mean{\lVert\boldw^{\AP}_{l,i}\rVert^2} \!+\! L (M\! P_{\dsp} \!+\! P_{\hov}) \!\leq\! t, \label{mee2_a} \\[-.4em]
    & \log_2 \groupele{1+1/\gamma_k}\dfrac{1}{t} \geq r_k, \label{mee2_b} \\[-.4em]
    & \SINR^{(\dwnl)}_k (\pwwb) \geq 1/\gamma_k. \label{mee2_c} 
\end{align}
\end{subequations}
The constraint \eqref{mee1_a} is linear in the variables $\{\pww^{\AP}_{l,i}\}.$ While, the positivity of the second derivative within the domain $\gamma_k \!>\! 0$ confirms the convexity of the constraint \eqref{mee1_b}. The constraint \eqref{mee2_a} is convex because it combines a quadratic function in the optimization variables with a term linear in $t$. Moreover, the constraint \eqref{mee2_a} is convex because it is the sum of squares of absolute values and vector norms, which are both convex operations. Assume that, at the $(n\!-\!1)$-th iteration, a feasible solution of the EEM problem includes $\gamma^{(n-1)}_k$ and $t^{(n-1)}.$ By using the first-order Taylor expansion, the constraint \eqref{mee2_b} can be approximated by a convex constraint as follows
\begin{align}
    &\dfrac{ \log_2 \big(1 \!+\! 1/\gamma^{(n-1)}_k \big)}{t^{(n-1)}} \!-\! \dfrac{ \log_2 \big(1 \!+\! 1/\gamma^{(n-1)}_k \big) (t \!-\! t^{(n-1)})}{(t^{(n-1)})^2 } \nonumber \\[-.5em]
    &\quad - \frac{ (\gamma_k - \gamma^{(n-1)}_k)}{{\Big(\gamma^{(n-1)}_k\Big)}^2 t^{(n-1)} \Big(1 + 1/\gamma^{(n-1)}_k \Big) \ln(2)} \! \geq \! r_k.
\end{align}
The non-convexity of constraint \eqref{mee2_c} can be addressed by using the auxiliary variables $\xib = \{\xi_i\}_{i=1}^K$ as follows
    \begin{numcases}{}
        \!\!\!\!\!\!\!&$\pww^{\SN}_k\numabs{\mean{\chg^\Hrm_k \boldw^\SN_k}}^2 + \numabs{\boldb^\Hrm_{k,k} \pwwb_k}^2 \geq \xi_k / \gamma_k, \quad \forall k \in \UESet$, \label{mee2_c1} \\
        \!\!\!\!\!\!\!&$\sum\limits_{i=1}^K \pww^{\SN}_i \mean{\numabs{\chg^\Hrm_k\boldw^{\SN}_i}^2} - \pww^{\SN}_k\numabs{\mean{\chg^\Hrm_{k}\boldw^{\SN}_k}}^2$ \nonumber \\[-.8em]
        \!\!\!\!\!\!\!&$\qquad + \sum\limits_{i=1}^K \vecnorm{\Cmat^\Hrm_{k,i}\pwwb_i}^2 + \sigma^2_{\dwnl} \leq \xi_k, \quad \forall k \in \UESet.$ \label{mee2_c2}
    \end{numcases}
We can observe that the constraint \eqref{mee2_c1} is non-convex, while the constraint \eqref{mee2_c2} is convex due to the quadratic forms $\vecnorm{\Cmat^\Hrm_{k,i}\pwwb_i}^2$. Assume that, at the $(t-1)$-th iteration, a feasible solution includes $\xi^{(t-1)}_k$ and $\gamma^{(t-1)}_k.$ The non-convex constraint \eqref{mee2_c1} can be addressed by using the first-order Taylor expansion in a neighborhood of $\big(\xi^{(t-1)}_k, \gamma^{(t-1)}_k\big)$ as follows: 
\begin{equation}\label{mee2_c1_Taylor}
    \arraycolsep=2pt\def\arraystretch{1.5}
    \begin{array}{l}
    \pww^{\SN}_k\numabs{\mean{\chg^\Hrm_k \boldw^\SN_k}}^2 + \numabs{\boldb^\Hrm_{k,k} \pwwb_k}^2 \geq {\xi^{(t-1)}_k}/{\gamma^{(t-1)}_k} \\
    + \dfrac{1}{\gamma^{(t-1)}_k} \big(\xi_k - \xi^{(t-1)}_k\big) - \dfrac{\xi^{(t-1)}_k}{\big(\gamma^{(t-1)}_k\big)^2} \big(\gamma_k - \gamma^{(t-1)}_k\big).
    \end{array}     
\end{equation}

\noindent Finally, the EEM problem \eqref{mee2} can be approximately solved via SCA method as described in Algorithm \ref{alg:EEM}.
\begin{algorithm}[!t]\small
\caption{Proposed SCA Algorithm for EEM Problem.}
\begin{algorithmic}[1]
\State \textbf{Initialization:} $n \gets 0;$ $\varepsilon \gets 0.001;$ 
\State $\pww^{\SN}_i, \forall i = 1,\ldots , K,$ according to the EPA scheme. $\pwwb^{(n)}$ is a feasible solution which is obtained by random search method\footnotemark;
\State $\gamma^{(n)}_k \!\gets\! (2^{\SE^{\min}_k}-1)^{-1};~\forall\,k,$
\State $\xi_k \!\gets\! \gamma_k \groupele{\pww^{\SN}_k\numabs{\mean{\chg^\Hrm_k \boldw^\SN_k}}^2 \!+\! \numabs{\boldb^\Hrm_{k,k} \pwwb_k}^2};~\forall\,k,$
\State $t^{(n)} \!\gets\! \dfrac{1}{\epsilon}\sum\limits^{L}_{l = 1} \sum\limits^K_{i = 1} \!\pww^{\AP}_{l,i}\, \mean{\lVert\boldw^{\AP}_{l,i}\rVert^2} \!+\! L (M P_{\dsp} \!+\! P_{\hov});$
\State $r^{(n)}_k \!\gets\! \log_2 \groupele{1+1/\gamma^{(n)}_k}/t^{(n)}$
\State $\boldr^{(n)} \!\gets\! \{r_1\ldots r_k\};$ $\gammab^{(n)}\!\gets\!\{\gamma_1\ldots \gamma_k\};$ $\xib^{(n)} \!\gets\! \{\xi_1\ldots \xi_k\};$ 
\Repeat
    \State Solve~\eqref{mee2} and let $\groupele{\pwwb^*,\boldr^*,\gammab^*,\xib^*,t^*}$ be its solution;
    \State $\groupele{\pwwb^{(n+1)}, \boldr^{(n+1)}, \gammab^{(n+1)}, \xib^{(n+1)}, t^{(n+1)}} \gets$
    \State \phantom{aaaaaaaaa} $\groupele{\pwwb^*, \boldr^*, \gammab^*, \xib^*, t^*};$
    \State $n\gets n+1;$
\Until $\numabs{\sum\nolimits_{k=1}^K r^{(n)}_k \!-\! \sum\nolimits_{k=1}^K r^{(n-1)}_k} / \sum\nolimits_{k=1}^K r^{(n-1)}_k \!\leq\! \varepsilon;$
\State Output: $\pwwb^{(n)}, \boldr^{(n)}, \gammab^{(n)}, \xib^{(n)}, t^{(n)};$
\end{algorithmic}
\label{alg:EEM}
\end{algorithm}
\footnotetext{The random search method assigns points (i.e., coefficients $\pww^{\AP}_{l,k}$) within the interval $[0, P^{\AP}_{\dwnl}]$. Initially, the interval is divided into $G$ equally-spaced points. For each UAV, the algorithm iterates by randomly selecting one of the $G$ points. This process runs until the constraints \eqref{mee1_a} and \eqref{mee1_b} are met.}

\section{Performance Evaluation} 
We consider an area of 16 km$^2$ wherein $K\!=\!40$ GUs are served by an integrated satellite-terrestrial UAV-enabled CF-mMIMO system with $L\!=\!60$ UAVs, each one equipped with $M\!=\!4$ antennas. The satellite is equipped with $N\!=\!100$ antennas and orbits at $550$ km, while the UAVs hover at $50$ m of altitude. 
System parameters include $P^{\AP}_{\dwnl} \!=\! 1$ W, $P_{\dsp} \!=\! 0.1$ W, $P_{\hov} \!=\! 50$ W, power amplifier efficiency $\epsilon \!=\! 0.8,$ angular standard deviation (ASD) $10^\circ,$ $d_0 \!=\! 1$ m reference path loss distance, $f_c = 6$ GHz carrier frequency, $1.2$ dB and $4$ dB noise figures at the UAVs and the satellite, respectively, $B\!=\!20$ MHz transmission bandwidth, and a thermal noise density of $-174 + 10 \log_{10}(B)$ dBm. The large-scale fading between $\AP_l$ and $\guser_k$, measured in [dB], is defined as $\beta_{l,k} \!=\! G_l + G_k -8.5 - 38.63 \log_{10}(d_{l,k}/d_0) - 20 \log_{10}(f_c) + z_{l,k}$, where the antenna gain at UAV $l$ is $G_l \!=\! 10$ dBi, the antenna gain at $\guser_k$ is $G_k \!=\! 10$ dBi, $d_{l,k}$ is the distance between $\AP_l$ and $\guser_k$,  and $z_{l,k} \!\sim\! \mathcal{CN}(0, \alpha^2)$ is the shadow fading with zero mean and standard deviation $\alpha \!=\! 6$ dB. The large-scale fading between the satellite and $\guser_k$ is given in [dB] as $\beta_{k} \!=\! G + G_k - 32.45 - 20 \log_{10}(d_k/d_0) - 20 \log_{10}(f_c) + z_k,$
where $G \!=\! 30$ dBi is the antenna gain at the satellite, $d_{k}$ is the distance between the satellite and $\guser_k$, and $z_k$ is the log-normal shadow fading with zero mean and its variance depending on the elevation angle, the channel condition, and the carrier frequency. 
The probability of having a LoS link \cite{Zeng8663615, Lee9217992} is given by
$\sbar{\prob} \!=\! [1 + a\exp(-b\theta+ab))]^{-1}$,
where $a\!=\!5$ and $b\!=\!0.05$ are environment-specific constants and $\theta$ is the elevation angle between the two nodes. The Rician factor for the SN-GU link is given by $\kappa^{\chg}_{k} \!=\! 9.5 + 10 \log_{10}(N) + 0.5 \log_{10}(d_k)$. The Rician factor for the UAV-GU link is given by $\kappa^{\chh}_{l,k} \!=\! 15 + 1.0 \log_{10}(d_{l,k}).$ The ULA's response, assuming half-wavelength antenna spacing, is
\begin{equation}
\bolda(\varphi) = \left [ 1, e^{j \boldk (\varphi)}, \ldots, e^{j (M-1) \boldk(\varphi)} \right ]^T \in \Cbb^M,
\end{equation}
where  $\boldk(\varphi) = \pi \sin(\varphi)$ is the wave vector with $\varphi$ denoting the AoA of the signal from a transmitter to a receiver. Lastly, we assume $\SE^{\min}_k \!=\! 0.2$ bit/s/Hz, $k\!=\!1,\ldots,K$.

As an alternative power allocation (PA) strategy to that established by the EE maximization (which we dubbed ``EEM PA''), we consider the \textit{fractional} PA (FPA), which consists in setting the power coefficients as
\begin{equation*}
    \pww^{\SN}_k \!=\! P^{\SN}_{\dwnl} \frac{[\trace{\Ecal^{\chg}_k}]^{\nu}}{\sum\limits^K_{i=1} [\trace{\Ecal^{\chg}_i}]^{\nu+1}}\,, \text{ and } \pww^{\AP}_{l,k} \!=\! P^{\AP}_{\dwnl} \frac{[\tr{\Ecal^{\chh}_{l,k}}]^{\nu}}{\sum\limits^K_{i=1} [\tr{\Ecal^{\chh}_{l,i}}]^{\nu+1}}\,, 
\end{equation*}
where $\nu$ determines the power allocation behavior. For instance, the equal power allocation (EPA) can be obtained from the FPA upon setting $\nu\!=\!-1$, with each UAV transmitting to each GU with power $\pww^{\AP}_{l,k}\,  \mean{\lVert\boldw^{\AP}_{l,k}\rVert^2} \!=\! P^{\AP}_{\dwnl}/K\,,\forall\,l,\forall\,k$. The coefficients $\{\pww^{\SN}_k\}$ are determined based on the EPA scheme whenever the EEM algorithm is used.%

\begin{figure}[!t]
    \centering
    \includegraphics[width=0.8\linewidth]{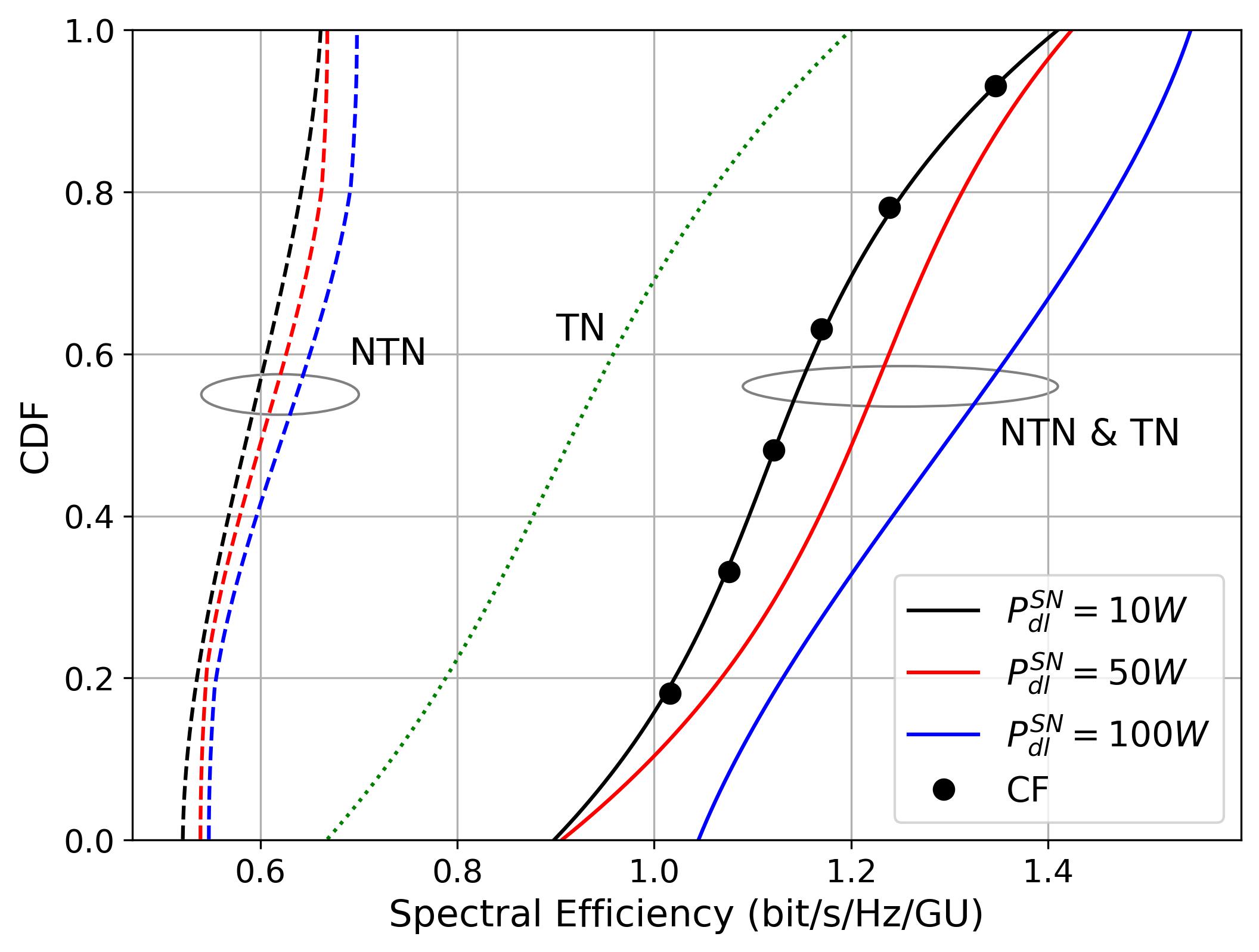}
    \vspace{-3mm}
    \caption{CDF of the per-GU SE for $L\!=\!60$, $K\!=\!40$ and $\SE^{\min}_k \!=\! 0.2$ [b/s/Hz], $\forall k$, under different $P^{\SN}_{\dwnl}$ values and network operations. Results are obtained via Monte Carlo simulations. Label ``CF'' refers to the results obtained by implementing the derived closed-form expressions for the SE.}
    \label{fig1}
    \vspace{-3mm}
\end{figure}
Fig.~\ref{fig1} shows the cumulative distribution function (CDF) of the per-GU SE over several network snapshots obtained from the numerical evaluation of the expectations via Monte Carlo  (MC) simulations and from our derived closed-form (CF) SE expressions. We consider different network operations: $(i)$ satellite and UAVs cooperating in a cell-free setup (i.e., NTN \& TN); $(ii)$ only UAVs operating in a pure terrestrial cell-free network (i.e., TN); and $(iii)$ only the satellite operating in an NTN. These results are obtained by implementing the EEM algorithm and considering three different transmit powers at the satellite, that is $P^{\SN}_{\dwnl} \!=\! \{10,\,50,\,100\}$ W.
Fig.~\ref{fig1} reveals that the SE is significantly improved when both satellite and UAVs actively cooperate. When only UAVs are active, the SE experiences a significant decrease, whereas it severely drops when the GUs are served only by the satellite. The perfect matching between the MC and CF curves validates our closed-form SE expressions. The GUs benefit from the proximity of the UAVs which complements the satellite's broad reach. However, it is clear that a pure NTN provides, in this case, a poor SE, regardless of the transmit power.

\begin{figure}
    \centering
    \includegraphics[width=0.8\linewidth]{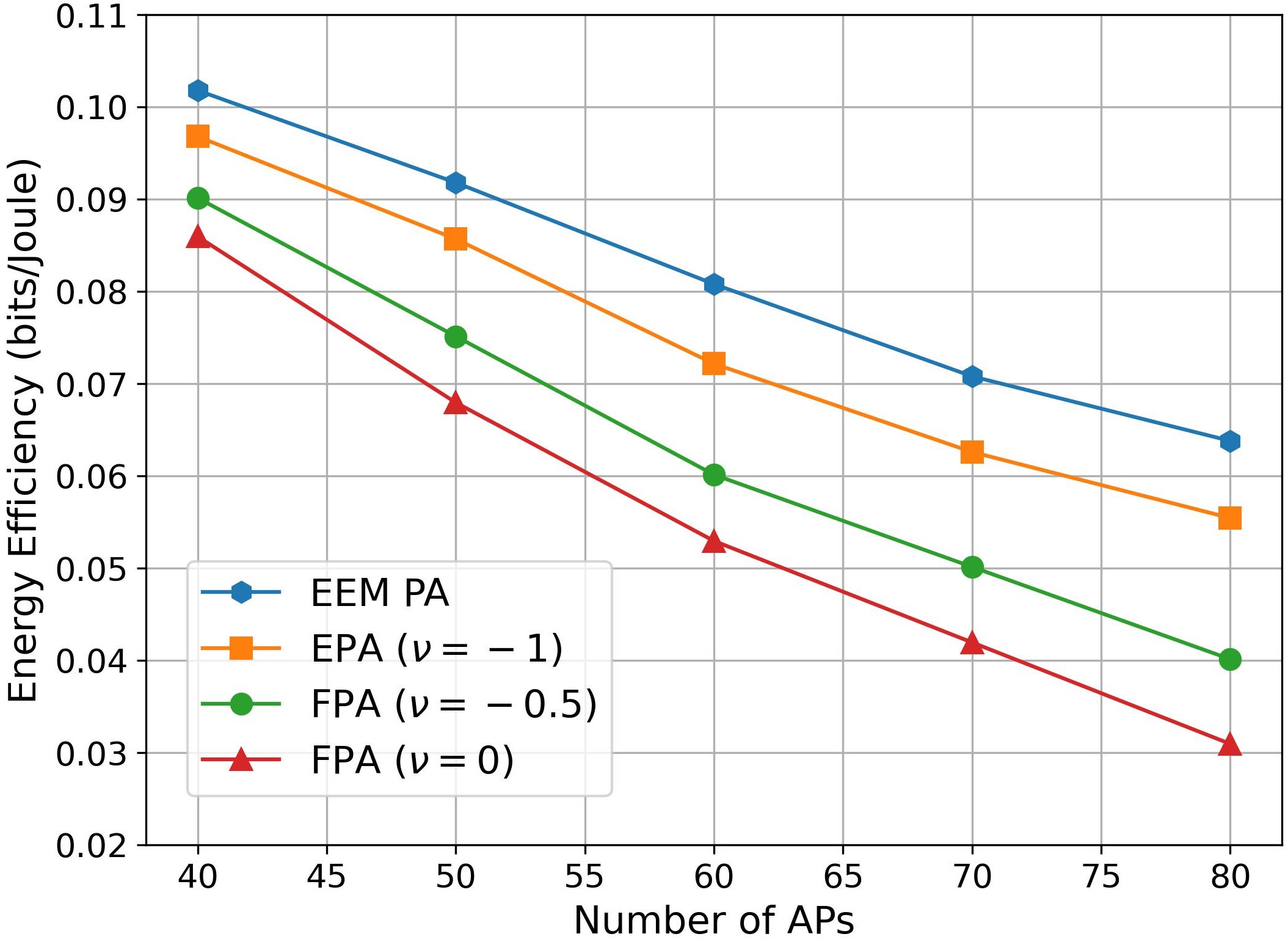}
    \vspace{-3mm}
    \caption{Average EE at the UAV layer achieved by the NTN \& TN system against the number of UAVs. We consider both FPA and EEM PA resulting from Algorithm~\ref{alg:EEM}. Here, $K=40$ GUs and $P^{\SN}_{\dwnl} = 10$ W.}
    \label{fig2}
\end{figure}
\begin{figure}
    \vspace{-2mm}
    \centering
    \includegraphics[width=0.8\linewidth]{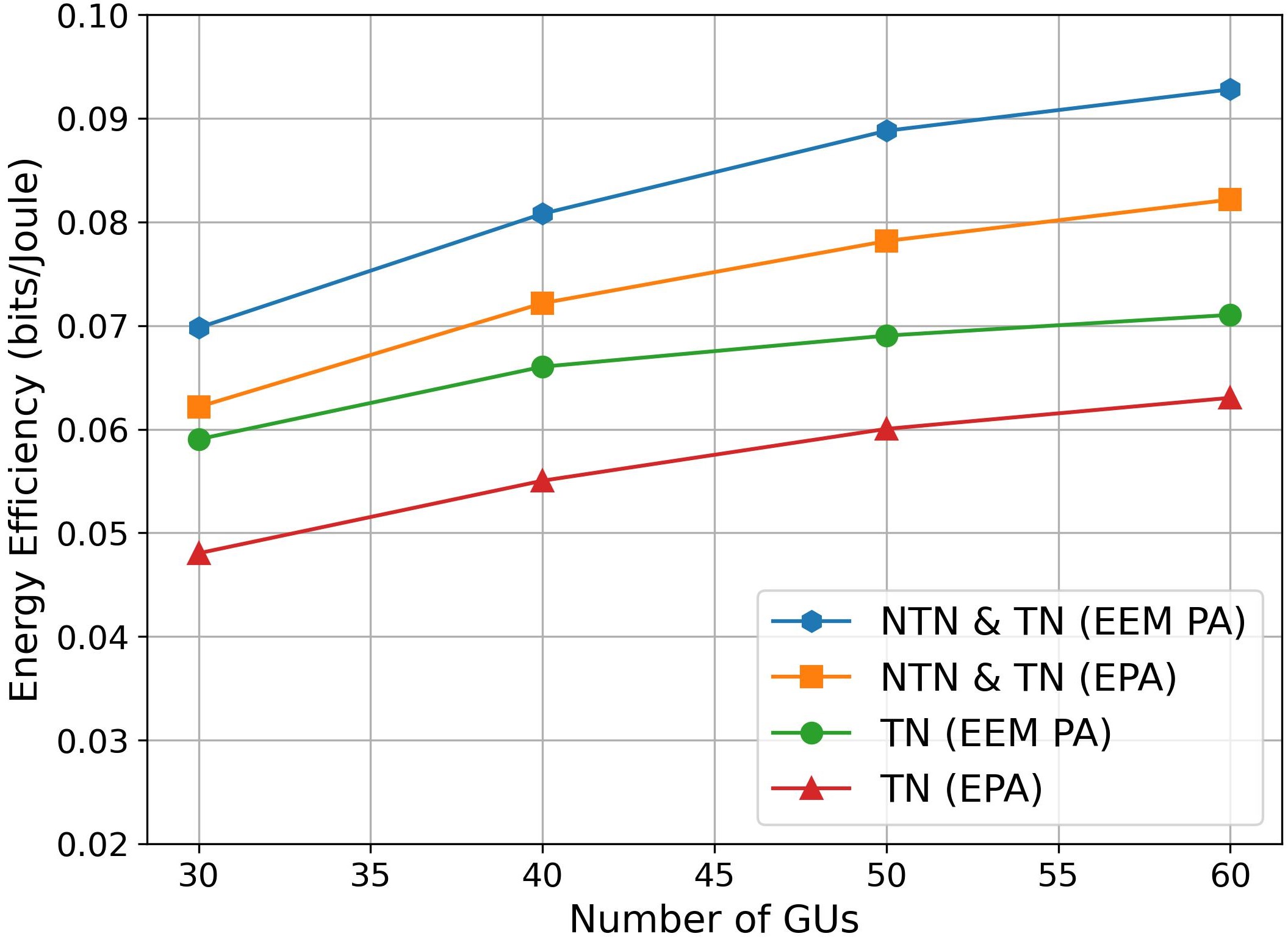}
    \vspace{-3mm}
    \caption{Average EE at the UAV layer against the number of GUs, under different network operations. We consider both EPA and EEM PA resulting from Algorithm~\ref{alg:EEM}. Here, $L=60$ UAVs and $P^{\SN}_{\dwnl} = 10$ W.}
    \label{fig3}
    \vspace{-3mm}
\end{figure}
Fig. \ref{fig2} illustrates the average EE at the UAV layer against the number of UAVs achieved by the integrated satellite-UAV CF-mMIMO system, with both FPA and EEM PA resulting from Algorithm~\ref{alg:EEM}, for $K\!=\!40$ and $P^{\SN}_{\dwnl} = 10$ W.
The EE at the UAV layer achieved by using Algorithm~\ref{alg:EEM} is higher than that achieved by employing FPA (including EPA), wherein no optimization is carried out. This gap demonstrates a real need of optimizing the transmit powers to achieve larger values of EE. Moreover, we observe that the EE uniformly decreases as the number of UAVs grows due to the increased energy costs. However, the EE gap between the EEM PA and FPA slightly increases, proving the effectiveness of our EEM algorithm.

Fig. \ref{fig3} compares the average EE at the UAV layer achieved by the aforementioned TN and NTN \& TN network operations against the number of GUs, considering both EEM PA and EPA (i.e., the FPA yielding the best performance in this scenario) and assuming $L\!=\!60$ UAVs and $P^{\SN}_{\dwnl} = 10$ W. 
The average EE increases with $K$ due to an increase of the sum SE. This suggests a great ability of the system in multiplexing the GUs despite the increased interference, especially when employing MR. The increase of EE is more pronounced in the integrated satellite-UAVs CF-mMIMO network.

\section{Conclusion}
In this work, we investigated the UAV-layer EE in the DL of an integrated satellite-terrestrial UAV-enabled CF-mMIMO network. In such a system, the satellite and the UAVs jointly and coherently cooperate to serve the GUs in a cell-free fashion.
We proposed an iterative algorithm based on the SCA method to solve the total EE maximization problem at the UAVs. The problem formulation relies on a closed-form expression of the DL EE that we derived assuming MR precoding. Simulation results reveal a real need of optimizing the UAV transmit powers to achieve larger values of EE as well as the effectiveness of the proposed EE maximization algorithm. Moreover, a few tens of UAVs transmitting with a fine-tuned power are sufficient to empower the service of the satellite and significantly increase the SE. On the other hand, an NTN additional link is able to boost the SE of a UAV-aided cell-free TN. 
Future extensions of this work may include the analysis of a modular architecture that integrates communication, mobile-edge computation~\cite{Interdonato2024MEC} and caching. 

\bibliographystyle{IEEEtran}
\bibliography{Refs}
\end{document}